# A NOVEL PYRAMIDAL-FSMN ARCHITECTURE WITH LATTICE-FREE MMI FOR SPEECH RECOGNITION

*Xuerui Yang, Jiwei Li, Xi Zhou*

Cloudwalk Technology, Shanghai, China.
{yangxuerui, lijiwei, zhouxi}@cloudwalk.cn

## ABSTRACT

Deep Feedforward Sequential Memory Network (DFSMN) has shown superior performance on speech recognition tasks. Based on this work, we propose a novel network architecture which introduces pyramidal memory structure to represent various context information in different layers. Additionally, res-CNN layers are added in the front to extract more sophisticated features as well. Together with lattice-free maximum mutual information (LF-MMI) and cross entropy (CE) joint training criteria, experimental results show that this approach achieves word error rates (WERs) of 3.62% and 10.89% respectively on Librispeech and LDC97S62 (Switchboard 300 hours) corpora. Furthermore, Recurrent neural network language model (RNNLM) rescoring is applied and a WER of 2.97% is obtained on Librispeech.

*Index Terms*— Automatic speech recognition, FSMN, lattice-free MMI, RNNLM

## 1. INTRODUCTION

Deep neural networks (DNN) have been applied as acoustic model (AM) on large vocabulary continuous speech recognition (LVCSR) system in recent years other than GMM-HMM models [1, 2]. Early works such as feedforward neural networks (FNN) [3] only takes current time steps as input. Recurrent neural network (RNN), especially long short-term memory network (LSTM), has demonstrated superior results in speech recognition tasks due to its cyclic connections [4] and utilization of sequential information. Convolutional neural network (CNN), in which local connectivity, weight sharing, and pooling techniques are applied, also outperforms previous works [8, 9].

However, the training of RNN relies on back-propagation through time (BPTT) [10], which may bring problems such as more time consuming, gradient vanishing and exploding [11] due to its complex computation. Teacher forcing or professor forcing [12] training can solve these problems in some degree, but reduces the robustness of RNN as well. Recently, a feedforward sequential memory network (FSMN) is proposed [13]. FSMN could model long-term relationship without any recurrent feedback. Moreover, to build very deep neural architecture, skip connection is applied to FSMN [14], which makes a large improvement to previous models. Meanwhile, time delay neural network (TDNN) and factorized TDNN (TDNN-F) [15] are also widely used feedforward networks.

Traditional DNN-HMM hybrid AM are trained on cross-entropy (CE) criterion. Since speech recognition is a sequential problem, several sequential discriminative training criteria are applied after CE training such as maximum mutual information (MMI) [16], minimum Bayes risk (MBR) [17] and minimum phone error (MPE) [18]. Inspired by the use of Connectionist Temporal Classification (CTC) in diverse recognition tasks [19, 20], a new method called lattice-free MMI [21] (LF-MMI/Chain model) is developed. This method could be used without any CE initialization; thus, less computation is allowed.

In this paper, we proposed a novel CNN Pyramidal-FSMN (pFSMN) architecture with LF-MMI and CE joint training. A pyramidal structure is applied in memory blocks. In this structure, bottom layers contain less context information while top layers contain more context information, which employs appropriate time dependency and reduces the number of parameters simultaneously. Besides, skip connections are added every several layer. Considering extracting more sophisticated features from original Mel-Frequency Cepstral Coefficients (MFCCs), CNN layers are deployed as the front-end.

In section 5, we evaluate the performance of this architecture on various speech recognition tasks. In the 300 hours Switchboard corpus, the proposed architecture achieves a current best word error rate (WER) of 10.89%. Further in the 1000 hours Librispeech corpus, the WER reaches 3.62%. In addition, RNN language model (RNNLM) has shown advances in decoding and rescoring in our experiments, in which above 1% absolute improvement is obtained compared with traditional N-gram language model.

## 2. FSMN

FSMN [13] is a feedforward fully connected neural network, appended with memory blocks in hidden layers as shown in Fig. 1(a). The memory block takes $t_1$ previous time steps and $t_2$ next time steps input into a fixed-length representation, then computes their block-sum as current output. Different from the original FSMN architecture, DFSMN [14] removes the direct forward connection and take memory block as the only input. To overcome the gradient vanishing and exploding problems, skip connection and the memory strides are introduced. Fig. 1(b) shows this structure.

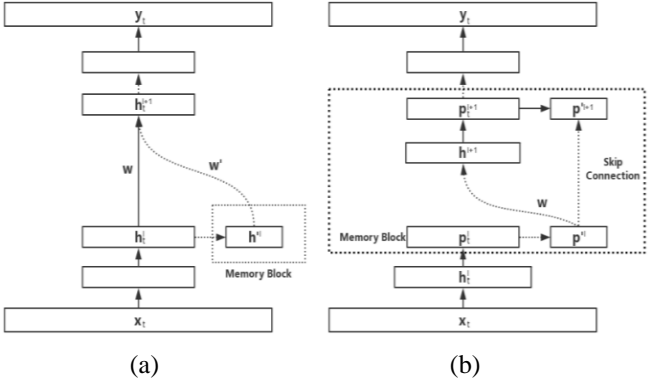

(a)                  (b)

**Fig. 1**. *FSMN(a) and DFSMN(b) architecture.*

The DFSMN component can be described as the following formulations:

$$h_t^{l+1} = f(W^l m_t^l + b_t^l) \tag{1}$$

$$m_t^{l+1} = m_t^l + h_t^{l+1} + \sum_{i=0}^{N_1^{l+1}} a_i^{l+1} \odot h_{t-s1*i}^{l+1} + \sum_{j=0}^{N_2^{l+1}} c_i^{l+1} \odot h_{t+s2*j}^{l+1} \tag{2}$$

$m_t^{l+1}$ denotes the output of the ($l$+1)-th memory block, and $h_t^{l+1}$ denotes the output of ReLU and linear layer. $a_i^{l+1}$ and $c_i^{l+1}$ are the coefficients in memory blocks. $s1$ and $s2$ are the strides.

## 3. LATTICE-FREE MMI

MMI aims to maximize the probability of the target sequence, while minimizing the probability of all possible sequences. For a training set $\mathbb{S} = \{(o^m, w^m)|0 \leq m \leq M\}$ where $o^m$ and $w^m$ denote the observed sequences and the correct sequence labels. The MMI criterion is:

$$\mathcal{J}_{MMI}(\theta; \mathbb{S}) = \sum_{m=1}^{M} \log \frac{p(o^m|w^m; \theta)^k p(w)}{\sum_w p(o^m|w'^m; \theta)^k p(w')} \tag{3}$$

where $p(w)$ represents the prior probability of word sequence **w** and $p(w')$ represents a feasible sequence in the search space.

Theoretically, the denominator requires all the word sequences but the computation would be too slow. Therefore, the practical denominator graph is estimated using lattices [22] generated by frame-level CE pre-training. More recently,

Povey et al. [21] use a lattice-free MMI in which an optimized hidden Markov model (HMM) topology motivated by CTC is adopted. In that method, 2-state left-to-right HMM is used which is similar to the phone-dependent blank labels of CTC. Then a forward-backward calculation on 4-gram phone language model is applied. Besides, AM in LF-MMI directly output pseudo log-likelihood instead of softmax output.

## 4. OUR APPROACH

### 4.1. Pyramidal structure

The memory block lengths are identical through all hidden layers both in FSMN and DFSMN. However, when bottom layers extract long context information at certain time step t, the bottom layers contain this as well. Thus, the long-term relationship is duplicated and no longer needed in top layers. In our approach, a pyramidal structure memory block is demonstrated, in which memory block extracts more context information with the layers go deeper. Hence, the bottom layers extract features on the phone-level information while top layers extract features on semantic-level and syntax-level information. This structure improves accuracy and reduces the number of parameters simultaneously.

With the modification of the memory block, skip connections of every layers will not be beneficial. Therefore, we reduce the number of skip connections. Only when there is a difference in memory block length, the skip connection is added. By optimizing the shortcuts, the gradients flow to deeper layers more efficiently. This could be formulated as follow, where m is the variation coefficients in memory blocks:

$$x_t^l = x_t^{l-m} + \sum_{i=0}^{N_1^l} a_i^l \odot h_{t-s1*i}^l + \sum_{j=0}^{N_2^l} b_i^l \odot h_{t+s2*j}^l \tag{4}$$

### 4.2. Residual CNN

Instead of directly using FSMN layers, a 6-layer CNN module is applied at the front end. The input data are reorganized into feature maps so that they could be fed into 2D-CNN. This is inspired by image-processing. In speech recognition tasks, the MFCC features and time steps correspond to pixel values of $x$ and $y$ respectively. With the network goes deeper, subsampling is applied to extract more robust features and reduce the features' size. Illustrated in Fig. 2, for every other layer, subsampling is used. In addition, residual structure is added correspondingly to solve the gradient problems.

### 4.3. LF-MMI and CE joint training

Sequential training usually tends to overfit. To avoid this problem, a CE loss output layer is deployed. The MMI and

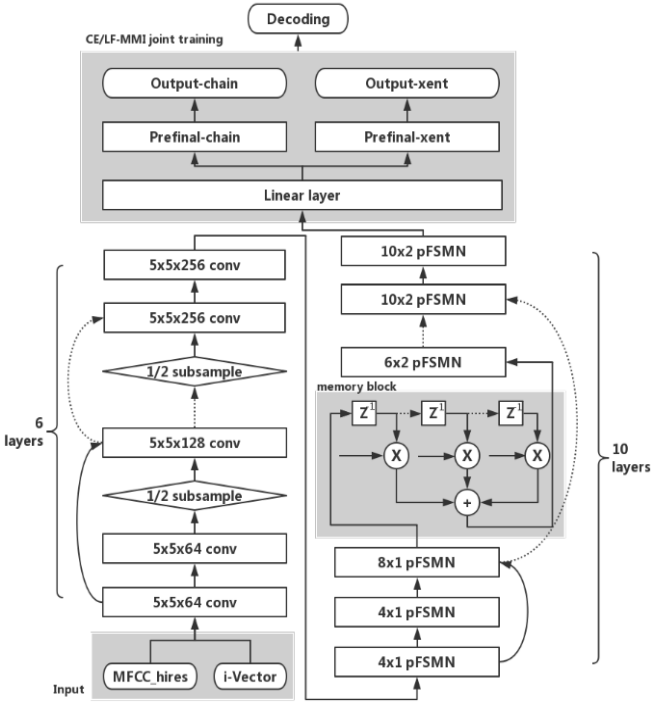

**Fig. 2**. *CNN-pFSMN AM architecture, in which memory block structure is demonstrated.*

CE criteria are combined with weighted sum, also known as CE regularization, seen in (5) and (6):

$$\mathcal{L}_{final}(\boldsymbol{x}) = \mathcal{L}_{LFMMI}(\boldsymbol{x}) + \alpha \mathcal{L}_{CE}(\boldsymbol{x}) \quad (5)$$

$$\mathcal{L}_{CE}(\boldsymbol{x}) = -\sum_{i=1}^{K} t_{pi} \log(y_i(\boldsymbol{x}_p)) \quad (6)$$

where $\alpha$ is the interpolation weight related with the CE regularization. $\mathcal{L}_{LFMMI}(\boldsymbol{x})$ is described in (3). During training, both LF-MMI and CE contribute to the parameters updating and loss computation. While decoding, only LF-MMI branch is used to generate accurate alignment for the network outputs.

### 4.4. RNNLM rescoring

During decoding, an N-gram LM is built for generating lattice and scoring. Even though N-gram LM is faster, it is still a statistic based LM and cannot utilize long context dependency. To achieve better results, we train a 5-layer TDNN-LSTM LM for rescoring. LSTM layers and TDNN layers appear alternately. And TDNN layers concatenate the current time step and several previous time steps. The initial decoding produces N-best word sequences; these sequences are then fed into LM network.

## 5. EXPERIMENTS

In this section, the performances of our approach are evaluated on different LVCSR English corpora including conversational and non-conversational speech. The results are compared with other popular models.

### 5.1. Experiment setup

The training data are 300 hours Switchboard corpus (SWBD-300) and 1000 hours Librispeech corpus. Five test sets are evaluated: test-clean, test-other, dev-clean, dev-other and train-dev seen in Table 1. For decoding, we train a 4-gram word LM and an RNNLM with 14500 books texts for Librispeech task and with Fisher+Switchboard texts for SWBD-300 task.

For running experiments, Kaldi [23] is used. HMM-GMM models are trained to generate force-alignment for neural network training. The features fed to all the evaluated models are 40-dimensinal MFCC and 100-dimensional i-vectors for speaker adaption. The MFCC features are extracted from 25 ms Hamming window for every 10 ms. We also adopt 3-fold speed perturbation (0.9x/1.0x/1.1x) for data augmentation in order to build more robust model.

The network architecture is shown in Fig. 2. 3*3 and 5*5 conv layers are employed at the front, and shortcuts are added when kernel size varies. 10 pFSMN blocks contain a block sum layer, a linear layer and a ReLU layer with dimensions of 1536, 256 and 256 in each block. Both CNN and pFSMN layers are penalized with L2 regularization. The time orders and strides are from 4 to 20 and 1 to 2 respectively. RNNLM is a 5-layer TDNN-LSTM network with 1024 dimension.

### 5.2. Results

Table 1 illustrates the WERs of various models. The results show that LF-MMI based method achieves obvious improvements than the previous CE based [14] method. Further, pFSMN with LF-MMI outperforms the previous models on both large, non-conversational (Librispeech) and small, conversational (SWBD-300) corpora. For example, in SWBD-300, WER of our approach drops from 11.15% to 10.89% compared with the BLSTM model. Moreover, by using RNNLM for n-best rescoring with n equals to 20, the WERs have achieved 2.97% in test-clean and 10.03% in train-dev respectively. This proves RNNLM has a superior performance than the conventional N-gram LM. However, it also slows down the decoding speed. Therefore, a trade-off needs to be considered when choosing between these two methods.

Various tricks are experimented as in Table 2. It presents that conv layers with shortcuts are beneficial to the model. And 5*5 kernel is relatively better among diverse kernel sizes. For memory block structure, the results show that big end pyramidal (top layers contain more context information) structure outperforms small end (bottom layers contain more context information) pyramidal structure. We have also tried semi-orthogonal constraints on linear bottleneck layer in [15], but it does no benefit to the model according to Table 2 1b and 1c.

**Table 1.** *Comparison of our approach and previous methods on Librispeech and SWBD-300 tasks.*

| Models | Librispeech | | | | SWBD |
|---|---|---|---|---|---|
| | test-clean | dev-clean | test-other | dev-other | train-dev |
| TDNN [24] | 4.17 | 3.87 | 10.62 | 10.22 | 13.91 |
| BLSTM [24] | - | - | - | - | 11.75 |
| TDNN-F [24] | 3.8 | 3.29 | 8.76 | 8.71 | 11.15 |
| DS2 [20] | 5.15 | - | 12.73 | - | - |
| ESPnet [25] | 4.6 | 4.5 | 13.7 | 13.0 | 17.6 |
| DFSMN [14] | 3.96 | 3.6 | 10.39 | 10.21 | - |
| DFSMN-Chain | 3.84 | 3.45 | 9.01 | 8.9 | 11.99 |
| CNN-pFSMN-Chain | **3.62** | **3.28** | **8.45** | **8.37** | **10.89** |
| +RNNLM rescoring* | **2.97** | **2.56** | **7.5** | **7.47** | **10.03** |

\* RNNLM rescoring is used on n-best sentences after decoding.
\* train-dev set is the first 4000 utterances from SWBD-300 data, the training set is the rest part.

**Table 2.** *WER gain of AM with various architectures on SWBD-300 task.*

| Models | Size | SWBD | | | |
|---|---|---|---|---|---|
| | | train-dev | Gain | +RNNLM rescoring | Gain |
| Baseline (DFSMN-chain) | 14M | 11.99 | - | 10.97 | - |
| 1a(c3*os1*) | 19M | 11.47 | +4.3% | 10.27 | +6.4% |
| 1b(c3s2o*) | 18M | 11.21 | +6.5% | 10.22 | +6.8% |
| 1c(c3s2) | 18M | 11.06 | +7.8% | 10.13 | +7.7% |
| 1d(c3s2op*) | 18M | 10.95 | +8.7% | 10.07 | +8.2% |
| 1f(c3s3p) | 18M | 11.33 | +5.5% | 10.23 | +6.7% |
| 1g(c5s2p) | 18M | **10.89** | +9.2% | **10.03** | +8.6% |

\*cn: conv layers with n*n kernels;
\*o: semi-orthogonal constraints;
\*sn: different skip connection method;
\*p: pyramidal-FSMN structure;

Fig. 3 (a) illustrates the training loss and accuracy of our final model. The model tends to converge after 300 iterations. Fig. 3 (b) is the WERs comparison of different decoding settings. The average improvement of RNNLM rescoring is more than 1%.

## 6. CONCLUSIONS

In summary, we proposed a novel CNN-pFSMN architecture trained with LF-MMI and rescored with RNNLM. By apply-

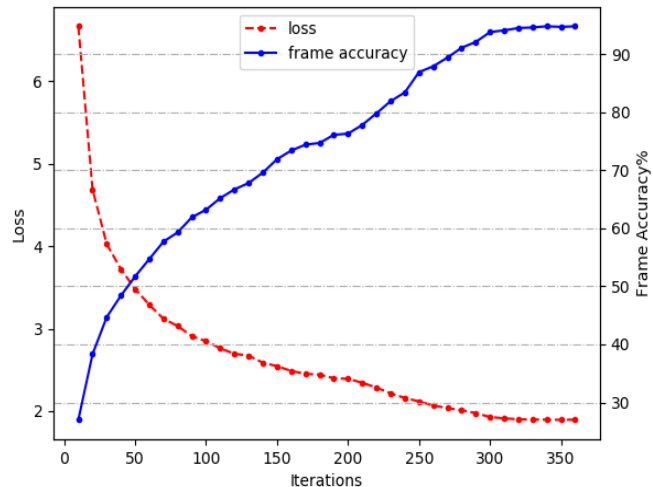

(a)

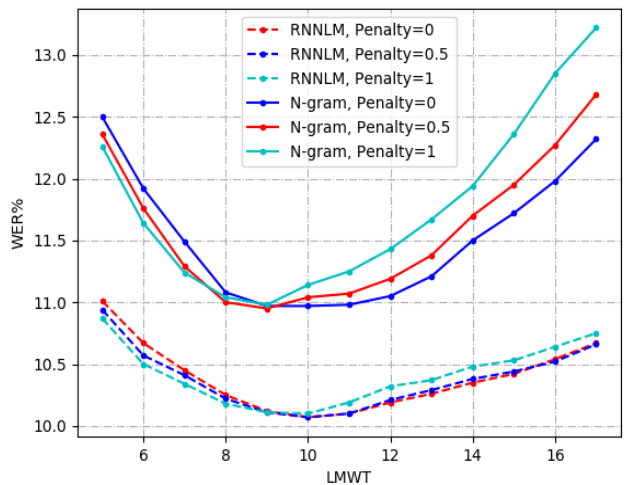

(b)

**Fig. 3.** *(a) The learning curves of various architecture on SWBD-300 task. (b) The WERs with different decoding settings on SWBD-300 task. LMWT is the language model weight.*

-ing res-CNN, CE/LF-MMI joint training and pyramidal memory structure, this approach has constantly outperformed the previous models on both Librispeech and Switchboard tasks. Other tricks like L2 regularization and speed perturbation are also beneficial. The results on test-clean and train-dev test set are 2.97% and 10.03% with rescoring respectively.

The experimental results prove that the pyramidal structure could extract phone-level, semantic-level and syntax-level information at different layers. Inspired by [26], the combination of LF-MMI and end-to-end on our architecture will be further explored in recent future.

## 6. REFERENCES

[1] Li Deng and Xiao Li, "Machine learning paradigms for speech recognition: An overview," *IEEE Transactions on Audio, Speech, and Language Processing*, vol. 21, no. 5, pp. 1060–1089, 2013.

[2] Geoffrey Hinton, Li Deng, Dong Yu, George E Dahl, Abdel-rahman Mohamed, Navdeep Jaitly, Andrew Senior, Vincent Vanhoucke, Patrick Nguyen, Tara N Sainath, et al., "Deep neural networks for acoustic modeling in speech recognition: The shared views of four research groups," *IEEE Signal processing magazine*, vol. 29, no. 6, pp. 82–97, 2012.

[3] George E Dahl, Dong Yu, Li Deng, and Alex Acero, "Context-dependent pre-trained deep neural networks for large-vocabulary speech recognition," *IEEE Transactions on audio, speech, and language processing*, vol. 20, no. 1, pp. 30–42, 2012.

[4] Tomas Mikolov, Martin Karafiat, Lukas Burget, Jan Cernocky, and Sanjeev Khudanpur, "Recurrent neural network based language model," in *Eleventh Annual Conference of the International Speech Communication Association*, 2010.

[5] Andrew W Senior, Hasim Sak, and Izhak Shafran, "Context dependent phone models for lstm rnn acoustic modelling.," in *ICASSP*, 2015, pp. 4585–4589.

[6] Hasim Sak, Andrew Senior, and Françoise Beaufays, "Long short-term memory recurrent neural network architectures for large scale acoustic modeling," in *Fifteenth annual conference of the international speech communication association*, 2014.

[7] Alex Graves, Abdel-rahman Mohamed, and Geoffrey Hinton, "Speech recognition with deep recurrent neural networks," in *Acoustics, speech and signal processing (icassp), 2013 ieee international conference on*. IEEE, 2013, pp. 6645–6649.

[8] Yanmin Qian, Mengxiao Bi, Tian Tan, and Kai Yu, "Very deep convolutional neural networks for noise robust speech recognition," *IEEE/ACM Transactions on Audio, Speech, and Language Processing*, vol. 24, no. 12, pp. 2263–2276, 2016.

[9] Ossama Abdel-Hamid, Abdel-rahman Mohamed, Hui Jiang, Li Deng, Gerald Penn, and Dong Yu, "Convolutional neural networks for speech recognition," *IEEE/ACM Transactions on audio, speech, and language processing*, vol. 22, no. 10, pp. 1533–1545, 2014.

[10] Paul J Werbos, "Backpropagation through time: what it does and how to do it," *Proceedings of the IEEE*, vol. 78, no. 10, pp. 1550–1560, 1990.

[11] Yoshua Bengio, Patrice Simard, and Paolo Frasconi, "Learning long-term dependencies with gradient descent is difficult," *IEEE transactions on neural networks*, vol. 5, no. 2, pp. 157–166, 1994.

[12] Alex M Lamb, Anirudh Goyal ALIAS PARTH GOYAL, Ying Zhang, Saizheng Zhang, Aaron C Courville, and Yoshua Bengio, "Professor forcing: A new algorithm for training recurrent networks," in *Advances In Neural Information Processing Systems*, 2016, pp. 4601–4609.

[13] Shiliang Zhang, Cong Liu, Hui Jiang, Si Wei, Lirong Dai, and Yu Hu, "Feedforward sequential memory networks: A new structure to learn long-term dependency," *arXiv preprint arXiv:1512.08301*, 2015.

[14] Shiliang Zhang, Ming Lei, Zhijie Yan, and Lirong Dai, "Deep-fsmn for large vocabulary continuous speech recognition," *arXiv preprint arXiv:1803.05030*, 2018.

[15] Daniel Povey, Gaofeng Cheng, Yiming Wang, Ke Li, Hainan Xu, Mahsa Yarmohamadi, and Sanjeev Khudanpur, " Semi-orthogonal low-rank matrix factorization for deep neural networks," *INTERSPEECH (2018, submitted)*, 2018.

[16] Lalit Bahl, Peter Brown, Peter De Souza, and Robert Mercer, "Maximum mutual information estimation of hidden markov model parameters for speech recognition," in *Acoustics, Speech, and Signal Processing, IEEE International Conference on ICASSP'86*. IEEE, 1986, vol. 11, pp. 49–52.

[17] Matthew Gibson and Thomas Hain, "Hypothesis spaces for minimum bayes risk training in large vocabulary speech recognition," in *Ninth International Conference on Spoken Language Processing*, 2006.

[18] Karel Vesely, Arnab Ghoshal, Lukas Burget, and Daniel Povey, "Sequence-discriminative training of deep neural networks.," in *Interspeech*, 2013, pp. 2345–2349.

[19] Alex Graves, Santiago Fernandez, Faustino Gomez, and Jurgen Schmidhuber, "Connectionist temporal classification: labelling unsegmented sequence data with recurrent neural networks," in *Proceedings of the 23rd international conference on Machine learning*. ACM, 2006, pp. 369–376.

[20] Dario Amodei, Sundaram Ananthanarayanan, Rishita Anubhai, Jingliang Bai, Eric Battenberg, Carl Case, Jared Casper, Bryan Catanzaro, Qiang Cheng, Guoliang Chen, et al., "Deep speech 2: End-to-end speech recognition in english and mandarin," in *International Conference on Machine Learning*, 2016, pp. 173–182.

[21] Daniel Povey, Vijayaditya Peddinti, Daniel Galvez, Pegah Ghahremani, Vimal Manohar, Xingyu Na, Yiming Wang, and Sanjeev Khudanpur, "Purely sequencetrained neural networks for asr based on lattice-free mmi.," in *Interspeech*, 2016, pp. 2751–2755.

[22] Philip C Woodland and Daniel Povey, "Large scale discriminative training of hidden markov models for speech recognition," *Computer Speech & Language*, vol. 16, no. 1, pp. 25–47, 2002.

[23] Daniel Povey, Arnab Ghoshal, Gilles Boulianne, Lukas Burget, Ondrej Glembek, Nagendra Goel, Mirko Hannemann, Petr Motlicek, Yanmin Qian, Petr Schwarz, et al., "The kaldi speech recognition toolkit," in *IEEE 2011 workshop on automatic speech recognition and understanding*. IEEE Signal Processing Society, 2011, number EPFL-CONF-192584.

[24] "Kaldi speech recognition toolkit," https://github.com/kaldiasr/kaldi.

[25] "Espnet: end-to-end speech processing toolkit," https://github.com/espnet/espnet.

[26] Albert Zeyer, Kazuki Irie, Ralf Schluter, and Hermann Ney, "Improved training of end-to-end attention models for speech recognition," *arXiv preprint arXiv:1805.03294*, 2018.